\documentclass[aps,pra,twocolumn,superscriptaddress]{revtex4-2}
\usepackage{graphicx}
\usepackage{amsmath}
\usepackage{textcomp}
\usepackage{hyperref}
\usepackage{color}

\begin{document}
\title{Phase transitions in a perovskite thin film studied by environmental in-situ heating nano-beam electron diffraction}

\author{Tobias Meyer}
\affiliation{4th Institute of Physics -- Solids and Nanostructures, University of Goettingen, Friedrich-Hund-Platz 1, 37077 G\"ottingen, Germany}

\author{Birte Kressdorf}
\author{Vladimir Roddatis}
\altaffiliation[Present address: ]{German Research Centre for Geosciences GFZ, Telegrafenberg, 14473 Potsdam, Germany}
\author{J\"org Hoffmann}
\author{Christian Jooss}
\affiliation{Institute of Materials Physics, University of Goettingen, Friedrich-Hund-Platz 1, 37077 G\"ottingen, Germany}

\author{Michael Seibt}
\email{mseibt@gwdg.de}
\affiliation{4th Institute of Physics -- Solids and Nanostructures, University of Goettingen, Friedrich-Hund-Platz 1, 37077 G\"ottingen, Germany}


\begin{abstract}
The rich phase diagram of bulk Pr$_{1-x}$Ca$_{x}$MnO$_3$ resulting in a high tunability of physical properties gave rise to various studies related to fundamental research as well as prospective applications of the material. 
Importantly, as a consequence of strong correlation effects, electronic and lattice degrees of freedom are vigorously coupled.
Hence, it is debatable whether such bulk phase diagrams can be transferred to inherently strained epitaxial thin films.
In this paper, the structural orthorhombic to pseudo-cubic transition for $x=0.1$ is studied in ion-beam sputtered thin films and point out differences to the respective bulk system by employing in-situ heating nano-beam electron diffraction to follow the temperature dependence of lattice constants. In addition, it is demonstrated that controlling the environment during heating, i.e. preventing oxygen loss, is crucial in order to avoid irreversible structural changes, which is expected to be a general problem of compounds containing volatile elements under non-equilibrium conditions.
\end{abstract}

\maketitle



Current and future technologies increasingly exploit the unique characteristics of metal oxide thin films,\textsuperscript{\cite{sangwan2020neuromorphic,park2020review,su2014silicon}}
where a plethora of physical properties can be tailored by temperature and composition dependent phase transitions.
Many of these compounds contain volatile components, i.e. elements in gaseous state at processing or even operating conditions. A prominent material class is metal oxides with well-known examples such as TiO$_x$ for improved solar cell contacts,\textsuperscript{\cite{hsu2015electron,guerrero2014light}} ZnO for ultraviolet light emitting diodes,\textsuperscript{\cite{yang2013largely,lupan2010low}} La$_{1-x}$Sr$_{x}$MnO$_{3}$ for solid oxide fuel cells,\textsuperscript{\cite{wang2017controlled,scholz2016rotating}} 
MgO for magnetic tunneling junctions,\textsuperscript{\cite{loong2014strain,oh2009bias,yuasa2004giant}} Li$_{1-x}$CoO$_{2}$ for solid state batteries,\textsuperscript{\cite{xue2018self,jung2013unexpected,luo2012binder}} or Pr$_{1-x}$Ca$_{x}$MnO$_{3}$ (PCMO) for neuromorphic computing.\textsuperscript{\cite{wan2019emerging,jeong2018nonvolatile,park2015electronic}} As a general feature, the oxygen activity adjusted by ambient conditions delicately controls properties and reaction paths in such systems. Referring to the metal oxides 
related to solid state batteries and neuromorphic computing,
such reaction paths and also applications rely on controllable, reversible phase transitions. Hence, typical non-equilibrium operation and processing conditions offer the route to steer the latter,\textsuperscript{\cite{zhang2018symmetry,jeen2013topotactic,janousch2007role}} but also 
involve the risk of undesired reaction paths.
In addition, the incorporation in structured devices raises the question whether and how phase diagrams obtained on bulk materials can be transferred to the nanomaterial counterparts.

Calcium doped PrMnO$_{3}$ (Pr$_{1-x}$Ca$_{x}$MnO$_{3}$, PCMO) is a prominent representative of strongly correlated manganites discussed in the context of neuromorphic computing, but also related to third generation photvoltaics\textsuperscript{\cite{raiser,ifland2,ifland1}} and catalysis.\textsuperscript{\cite{lole2020,mierwaldt2017environmental,mildnerETEM,raabe2012situ}}
Due to strong correlations, the system -- as many other strongly correlated materials\textsuperscript{\cite{tokura2000orbital}} -- has a rich phase diagram with several ordered phases showing remarkable properties.\textsuperscript{\cite{beaud2014time,sheng2012dynamics,cmr}}
Various studies have focused on magnetic ordering at low temperature which is well understood.\textsuperscript{\cite{Pollert,Jirak1,Jirak2}} In addition, phase transitions at higher temperature of several hundred degrees Celsius have been revealed by neutron and X-ray diffraction experiments on PCMO powders for low Ca concentrations.\textsuperscript{\cite{Pollert,Sanchez}} They mainly show the transition from an orthorhombic to a pseudo-cubic phase with small differences in lattice parameters. These changes are related to a reduction of the cooperative Jahn-Teller (JT) distortion as well as the alternating tilt of neighbouring MnO$_6$ octahedra in the high-temperature phase, leading to a higher isotropy. Still, direct evidence of orbital ordering driving the phase transition is lacking.

A major unsolved problem, however, is whether and how such data can be transferred to epitaxial thin films which are the typical basis for technological application. More generally, this is equivalent to the question whether bulk phase diagrams remain unchanged in thin film systems.
In fact, particularly in the context of transition metal oxides, various reports exist about tunable physical properties by epitaxial strain and oxygen vacancies possibly affecting phase transition temperatures as well.\textsuperscript{\cite{prop3,yang2019oxygen,enriquez2017oxygen,prop2}}

In this work, we investigate phase transitions at high temperature in PCMO ($x=0.1$) grown as a 400~nm thick epitaxial film on a SrTiO$_3$ (STO) substrate. In order to extract temperature-dependent lattice parameters we use nano-beam electron diffraction (NBED) in a 4D-STEM mode.\textsuperscript{\cite{LookInOphus}} Hence, individual domains of the typically nanotwinned films can be addressed in contrast to macroscopic X-ray or neutron diffraction studies. Furthermore, during the in-situ heating experiments, the environmental capabilities of the electron microscope are used to compare reaction paths in ultra-high vacuum (UHV) and 10\,Pa oxygen partial pressure ambient conditions. In the latter case a \textit{reversible} orthorhombic to pseudo-cubic phase transition is observed while in the former case an \textit{irreversible} phase transition associated with the appearance of superstructure reflections occurs, which completely suppresses the orthorhombic to pseudo-cubic phase transition. These results will be discussed in terms of oxygen vacancies whose formation is controlled by the ambient conditions.


\begin{figure}
    \centering
\includegraphics[width=.5\textwidth]{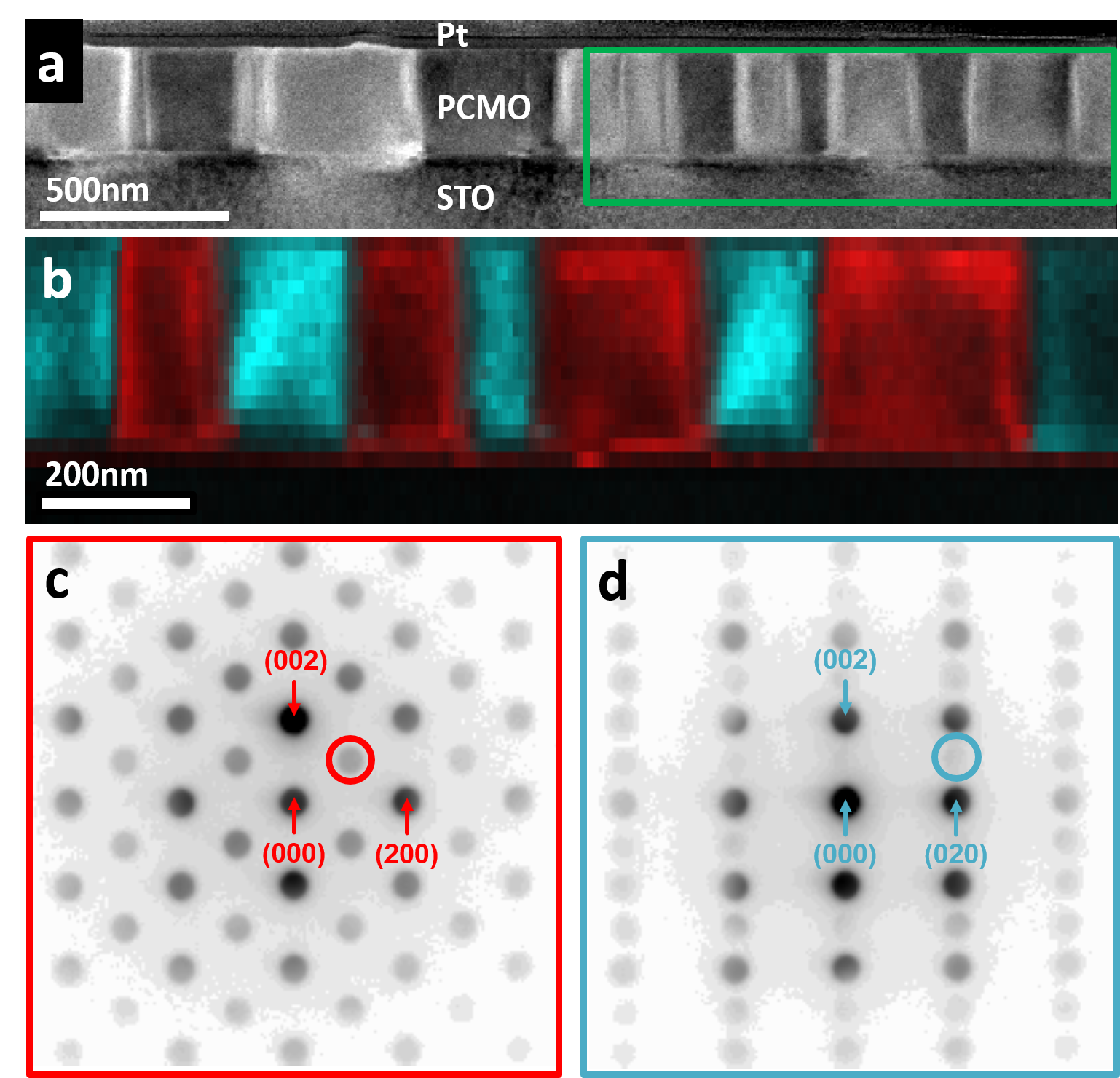}
    \caption{(a) ADF-STEM overview at room temperature with the scanning area marked by the green rectangle. (b) Domain structure of the PCMO thin film visualized by virtual dark-field images created by integrating the intensity of the PCMO (101) (red) resp. (021) (cyan) reflection. The [110] oriented STO substrate appears dark since neither of these reflections is allowed. (c) and (d) NBED patterns corresponding the red resp. cyan parts of the PCMO film with [010] resp. [100] zone axis orientation.}
    \label{fig:domains}
\end{figure}

\begin{figure}
    \centering
\includegraphics[width=.5\textwidth]{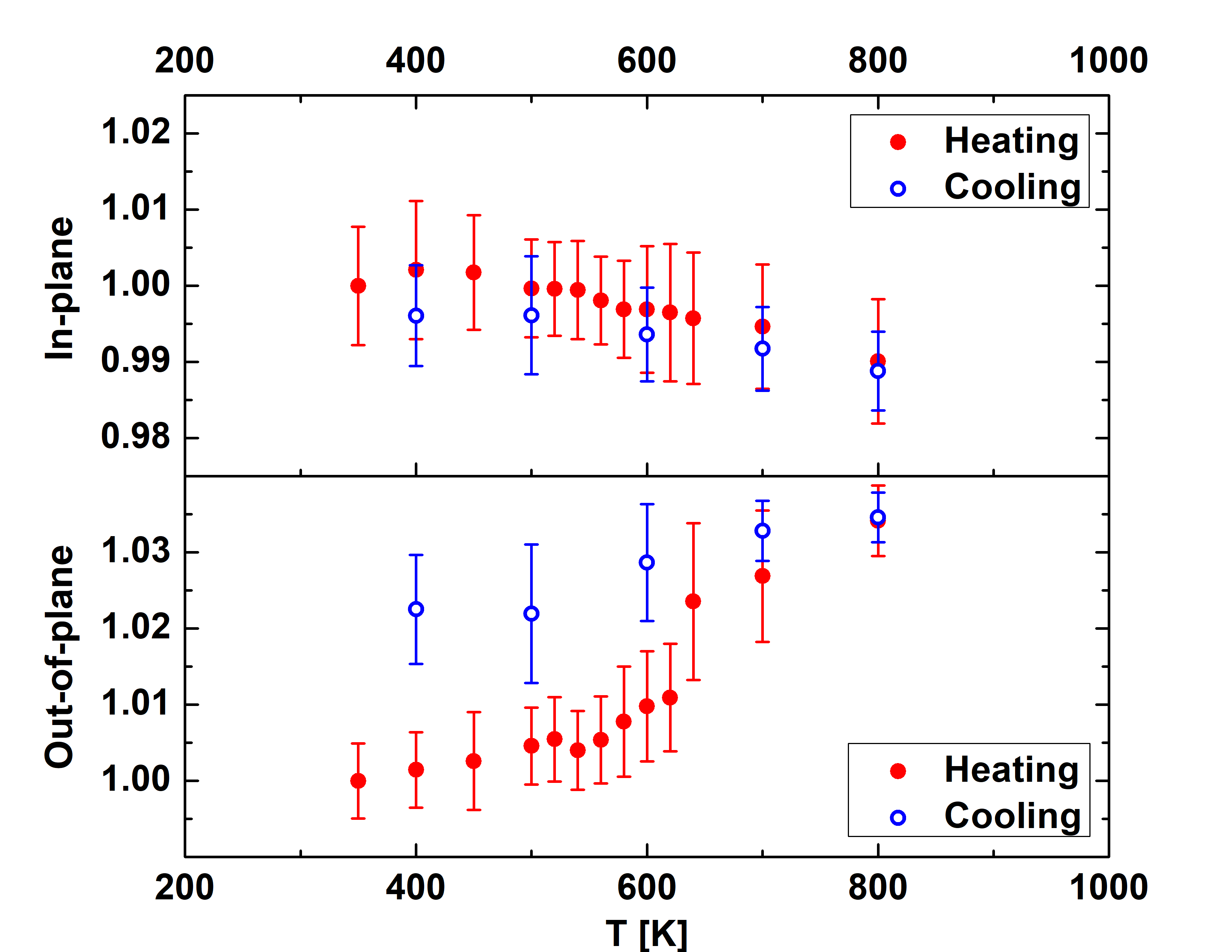}
    \caption{Temperature dependent in- and out-of-plane lattice parameters obtained under UHV conditions averaged over the upper half of the scanning range shown in Figure \ref{fig:domains} and normalized to their values at 350\,K. The error bars correspond to the standard deviations in the averaged range. 
    While the averaged in-plane parameter decreases slightly in temperature and exhibits a weak hysteresis, $c$ shows a strong increase as well as obvious hysteretic behavior.}
    \label{fig:HV}
\end{figure}

Prior to considering the temperature dependence of the given lattice parameters, a brief overview of the thin film epitaxy and heterogeneity shall be given. Generally, epitaxial growth throughout all investigated areas can be confirmed. As shown in Figure \ref{fig:domains},
the two relations [100]$_\text{PCMO}\|$[110]$_\text{STO}$ and [010]$_\text{PCMO}\|$[110]$_\text{STO}$ are predominantly found, i.e. [001]$_\text{PCMO}$ is oriented out-of-plane in both cases.
In fact, the in-plane orientation flips quite regularly on a few hundred nanometer scale and the resulting domains extend through the entire film as clearly visualized by the 4D-STEM analysis shown in Figure \ref{fig:domains}(b)-(d). 
This reflects a nanotwinned state of the film as it is often observed in orthorhombic systems.\textsuperscript{\cite{orthorhombictwins,orthorhombictwins2}}
Interestingly, there is no unique relation between orientation and bright and dark appearance in the ADF-STEM image in Figure \ref{fig:domains}(a), i.e. the two most left positioned [100] oriented (cyan) domains in Figure \ref{fig:domains}(b) appear bright resp. dark. This matches the observation of slightly varying zones axis alignments in the domains, i.e. a twisting, which does not solely correlate with the orientation but rather seems to be caused by the local strain state. Additionally, contrast heterogeneities in the STO are visible in Figure \ref{fig:domains}(a) as well indicating a strained lattice in the substrate. This is important since it prevents finding a (sufficiently thin) unstrained reference inside the TEM lamella and thus only relative values of lattice parameters will be considered.

\begin{figure}
    \centering
    \includegraphics[width=.49\textwidth]{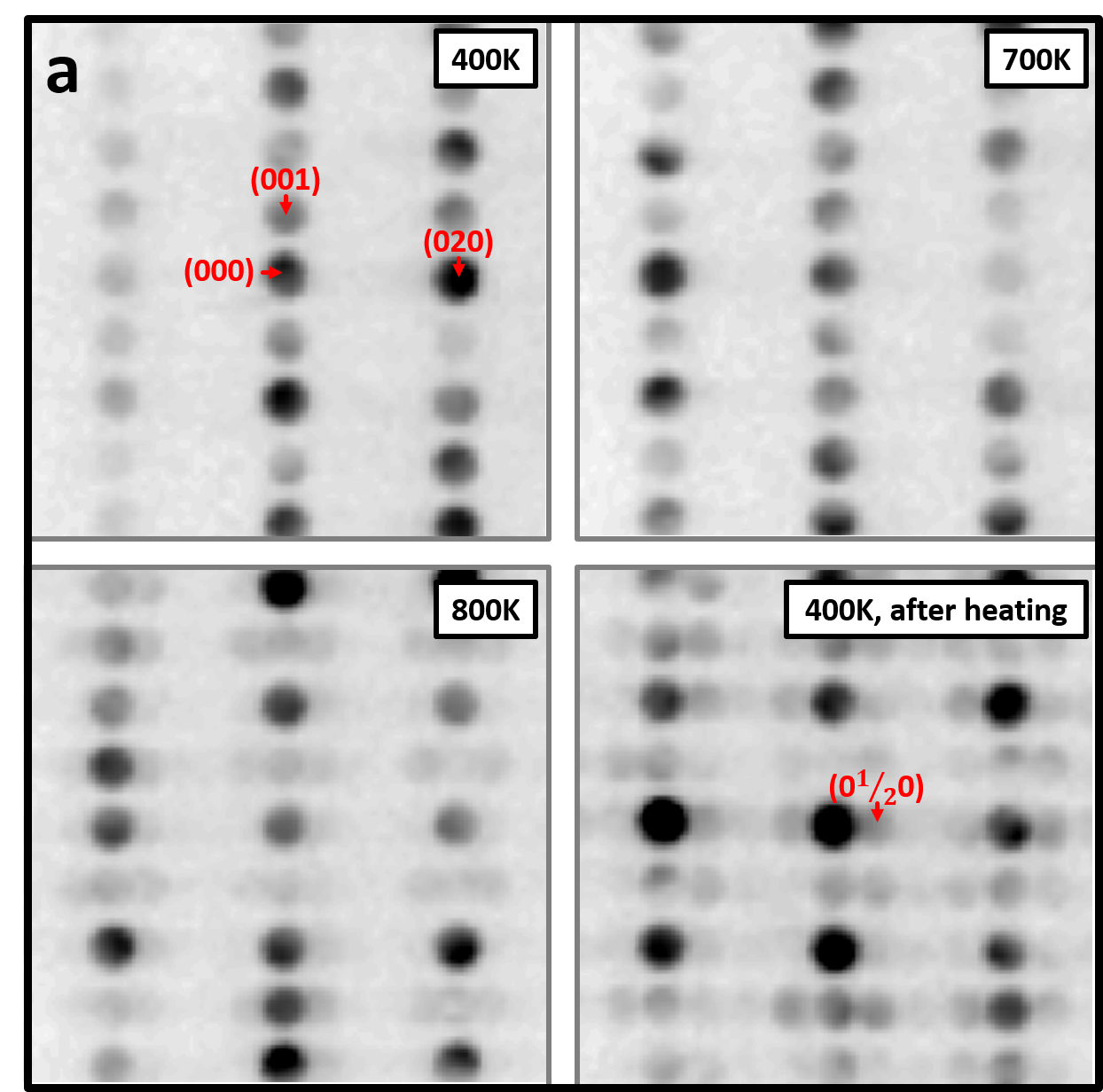}
    \includegraphics[width=.49\textwidth]{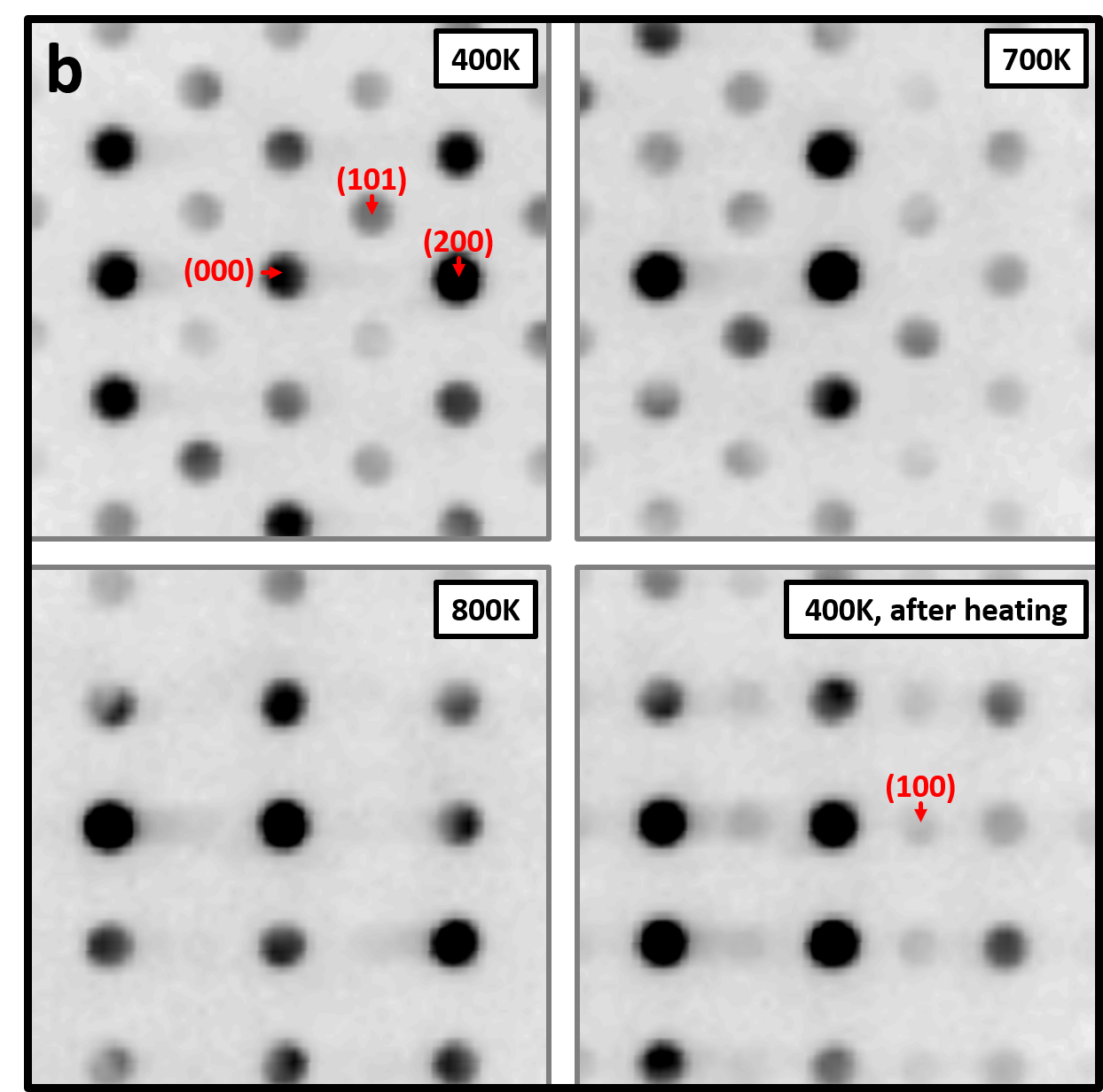}
    \caption{Temperature series of NBED patterns obtained under UHV conditions. (a) In [100] orientated domains, a $2b$ superstructure is observed first at $T=800$\,K which becomes increasingly obvious in time and persists at low temperatures after heating.
    (b) In [010] oriented domains, the (101) reflection vanishes at 800\,K and during the course of the experiment the initially extinct (100) reflections emerge.}
    \label{fig:SuperStructure}
\end{figure}

The temperature dependent lattice parameters averaged over the upper half of the scanning field in Figure \ref{fig:domains}
and normalized to their values at $T=350$\,K are shown in Figure \ref{fig:HV}. The error bars represent the standard deviation in the averaged range.
The in-plane parameter, including contributions from $a$ and $b$, decreases with higher temperature and shows a slight hysteresis.
In contrast, the out-of-plane parameter, i.e. $c$, increases and is strongly hysteretic. The latter indicates an irreversible change of the TEM lamella during heating which is confirmed by a change in the crystal symmetry: Figure \ref{fig:SuperStructure}(a) resp. (b) show series of NBED patterns of $a$- resp. $b$-oriented domains.
While in the former case a $2b$ superstructure emerges at 800\,K, the (101) reflections vanish in the latter.
Both effects persist at lower temperatures after heating and in addition, the initially extinct (100) reflections show up in $b$-oriented domains. 
The appearance of superstructures, although not exclusively occurring in $b$ direction, has been reported before in the context of catalytical studies of PCMO ($x=0.36$) and was attributed to oxygen vacancy ordering.\textsuperscript{\cite{mildnerETEM}} The circumstance that the effect is only observed in $b$ direction might be related to the results in Ref. \cite{ordering} for $x=1$ showing that in the presence of strain certain lattice sites are preferred for oxygen vacancy formation. Furthermore, the extinction rules are strongly coupled to the tilt pattern of neighbouring MnO$_6$ octahedra,\textsuperscript{\cite{Woodward}} and thus the observed changes in $b$-oriented domains suggest severe modifications in the collaborative tilt resp. distortion pattern.

\begin{figure}
    \centering
\includegraphics[width=.5\textwidth]{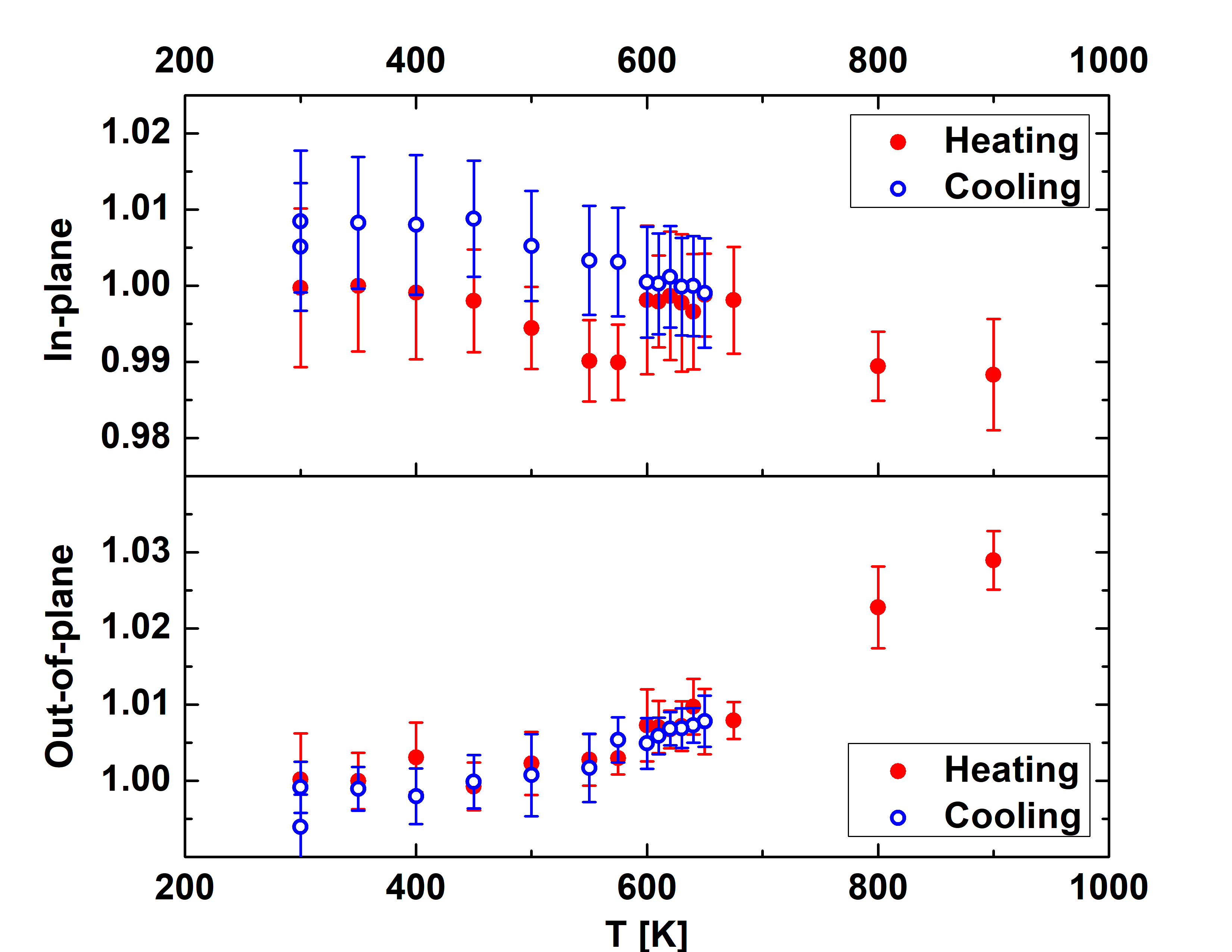}
    \caption{Temperature dependent averaged in- and out-of-plane lattice parameters obtained in a 10\,Pa oxygen environment normalized to their values at 350\,K. The error bars correspond to the standard deviations in the averaged range. 
    While the averaged in-plane parameter decreases in temperature and exhibits a sudden jump during heating at 600\,K, $c$ shows a strong increase and no hysteresis.
    }
    \label{fig:O2}
\end{figure}

\begin{figure}
    \centering
\includegraphics[width=.5\textwidth]{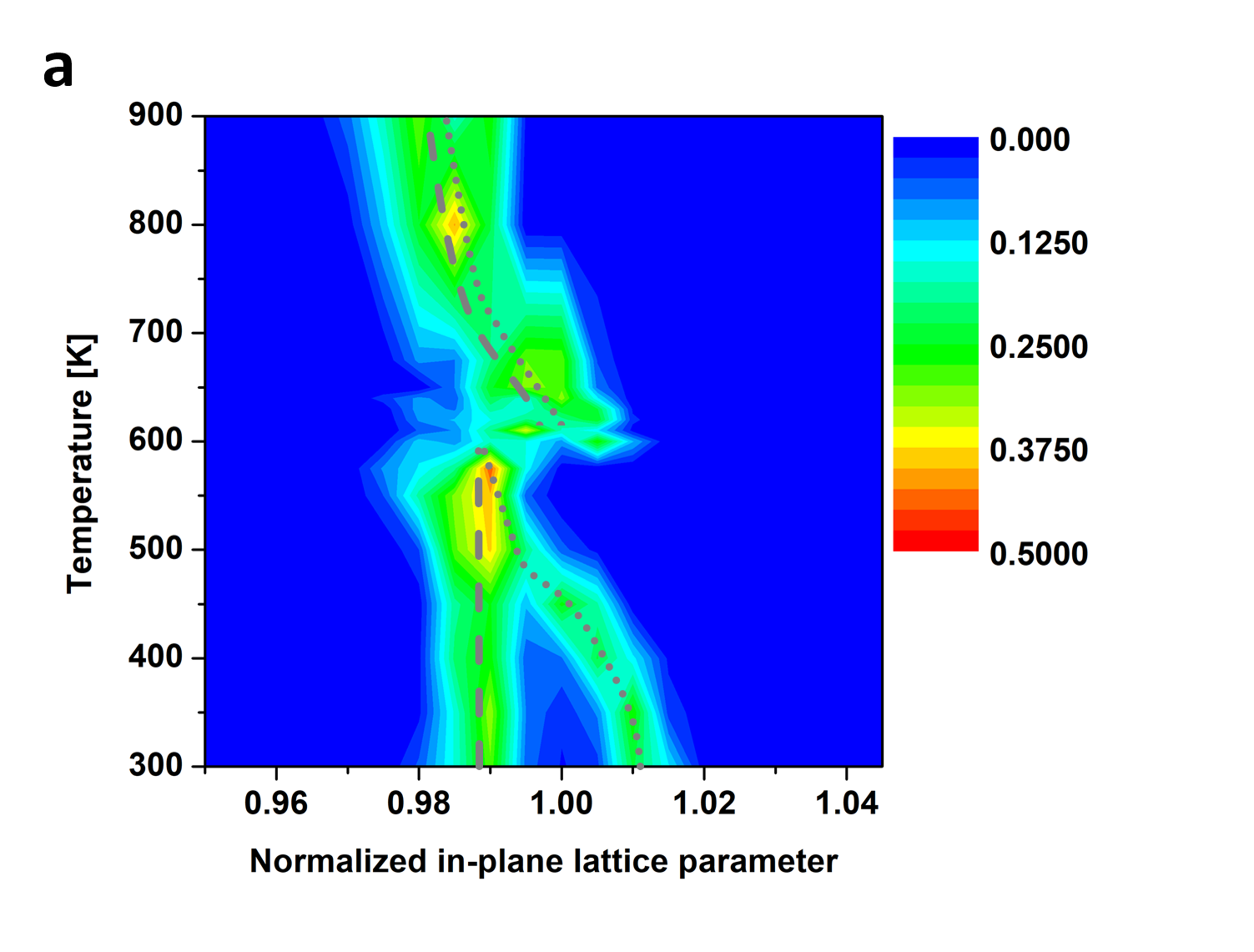}
\includegraphics[width=.5\textwidth]{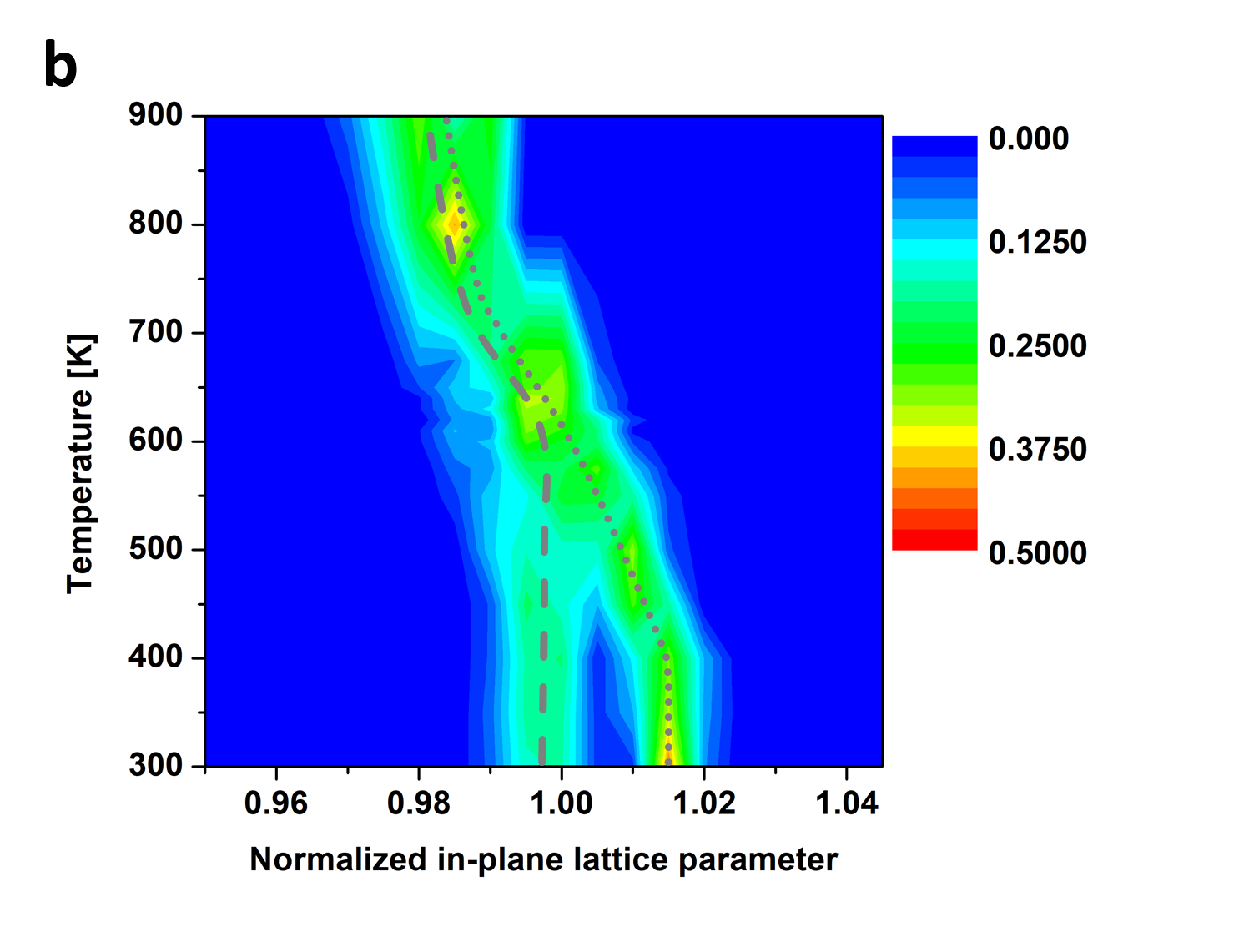}
    \caption{Contour plot of the temperature dependent histograms of normalized in-plane lattice parameters obtained in a 10\,Pa oxygen environment. The colors represent the relative frequency and the dashed resp. dotted grey lines are a guide to the eye to follow the $a$ resp. $b$ peak. (a) includes the data recorded during heating and shows a sudden jump at 600\,K right after the $a$- (small) and $b$-peaks (large) merge. (b) shows the data recorded during cooling (except for $T>650$\,K as only heating points exist) in which no jump is present. The splitting of $a$ and $b$ recurs after heating.}
    \label{fig:histograms}
\end{figure}

In order to suppress oxygen vacancy formation, an analogous experiment was conducted facilitating the ETEM gas-inlet capabilities and using an oxygen partial pressure of 10\,Pa.
The resulting lattice parameters obtained in the gaseous environment are shown in Figure \ref{fig:O2}.
As a matter of fact, $c$ still increases strongly with temperature but in contrast to the experiment in high-vacuum no hysteresis is visible.
Furthermore, neither a $2b$ superstructure nor a change in the extinction rules was observed in the gaseous environment confirming oxygen vacancy formation to be directly linked to the previous irreversible changes.
Nevertheless, the averaged in-plane parameter exhibits a sudden jump during heating at 600\,K which is not reverted during cooling and consequently the experiment is not entirely hysteresis-free.
This finding might be explained by thin sample effects, e.g. a lower energy strain state of the above mentioned twisting of neighbouring domains being possible due to the enhanced surface volume fraction in the TEM lamella. Alternatively, 
the extended annealing during the heating experiment could have lead to further strain relaxation, which is possible in macroscopic samples as well. Following the curve after the jump reveals 
a reversible decrease of the averaged in-plane parameter at high temperatures.
To disentangle the behavior of the so far averaged in-plane lattice parameters $a$ and $b$, the
temperature dependent histograms of detected in-plane lattice constants during heating and cooling are shown in Figure \ref{fig:histograms}(a) and (b), respectively. Please note that the histograms above $T=650$\,K in Figure \ref{fig:histograms}(b) are replotted from (a).
Generally, before and after heating, a bimodal distribution is observed at low temperatures and the 
separable peaks with lower resp. higher lattice parameter correspond to domains with [010] resp. [100] zone axis orientation.
As the temperature is raised, the two peaks start to merge and a -- possibly tetragonally strained -- pseudo-cubic phase forms.
Interestingly, the sudden jump during heating at 600\,K mentioned above occurs right at the pseudo-cubic transition in Figure \ref{fig:histograms}(a) suggesting that isotropic lattice constants are beneficial for this type of rearrangement.

Finally, the results obtained in oxygen environment after the rearrangement shall be compared to bulk measurements.
As described in Ref. \cite{Jirak1}, a pseudo-cubic transition is expected in the bulk for $x=0.1$ at approximately 675\,K.
Our findings show that the transition occurs at a slightly lower temperature in the thin film, i.e. at approximately 600\,K. 
Unfortunately, since no temperature dependent lattice constants are published at this calcium concentration, we can compare the trends only qualitatively to the existing data obtained in the bulk for $x=0$.\textsuperscript{\cite{Pollert,Sanchez}}
Here, due to the pseudo-cubic transition at 1050\,K, $c$ increases by 3.5\,\%, 
$b$ decreases by 3.5\,\%, and $a$ increases by 1.5\,\%. Since the initial anisotropy at room temperature is weaker for $x=0.1$, it is to be expected that theses changes are smaller due to the increased calcium content matching very well the observed increase of $c$ by 3.0\,\%
in Figure \ref{fig:O2} and the decrease of $b$ by 3.0\,\% in Figure \ref{fig:histograms}(b).
On the contrary, $a$ behaves qualitatively differently as it decreases with temperature.
Even though no direct comparison to the bulk can be drawn for $x=0.1$, it seems very likely that this effect is rather caused by strain than by the finite value of $x$. We justify this claim with the fact that the increase of $a$ is a fingerprint of the reduced MnO$_6$ tilting angle having a stronger effect on $a$ than the reduction of its JT distortion (which would in turn lead to an expansion in this direction). As the JT distortion is almost 50\,\% weaker for $x=0.1$ at room temperature\textsuperscript{\cite{Jirak1}} and the tolerance factor, i.e. the main driving force for octahedral tilts changes only by a few percent,\textsuperscript{\cite{Pollert}} we do not expect a reverted behavior due to the increased calcium content and attribute the reverted trend of $a$ to anisotropic strain.


In summary, nano-beam electron diffraction in combination with in-situ heating in an environmental transmission electron microscope was used to study phase transitions in PCMO thin films epitaxially grown on SrTiO$_3$ in the low-doping regime, i.e. for a Ca concentration of 10\,at\%. The oxygen activity was varied by the oxygen partial pressure during the in-situ experiment leading to two distinctly different routes followed by the system. 

On the one hand, UHV conditions accompanied by oxygen loss and hence formation of oxygen vacancies in the thin film material. From the appearance of a $2b$ superstructure accompanied by a change in the extinction rules as shown in Figure \ref{fig:HV} and \ref{fig:SuperStructure}, vacancy ordering can be concluded which persists after cooling to room temperature indicating a thermodynamically stable oxygen-deficient PCMO. We note here, that the observed anisotropy might be related to misfit strain resulting in an anisotropic oxygen vacancy formation enthalpy.

On the other hand, 10\,Pa oxygen partial pressure in the gaseous ambient inside the microscope column sufficiently reduces the oxygen loss and hence completely suppresses the hysteretic behavior of $c$, the change in extinction rules, as well as the superstructure formation. As a consequence, the clear detection of the reversible orthorhombic to pseudo-cubic phase transformation is possible with a slightly decreased critical temperature of approximately 600\,K when compared to the bulk counterpart.\textsuperscript{\cite{Jirak1}} In addition, a reverted behavior of the in-plane lattice parameter $a$ was observed which was attributed to
strain due to the
cubic substrate and the mosaic-like microstructure of the films. It is quite interesting to note, that the subtle irreversible relaxation occuring at about 600\,K during the first heating ramp coincides with the pseudo-cubic transition shown in Figure \ref{fig:histograms}(a), i.e. when $a$ and $b$ are approaching lifting the strain relaxation due to the alternating twinned domains. 

Finally, let us mention some implications of the results reported here. First of all, the study of phase transitions in thin film transition metal oxide perovskites by means of in-situ electron nanobeam diffraction in 4D-STEM mode is an approach with several advantages over bulk diffraction studies. We have clearly demonstrated that controlling the oxygen activity by ambient conditions is not only a pre-requisite for reliable in-situ heating diffraction studies, but also possible in an environmental transmission electron microscope. This important conclusion can be generalized to other systems containing volatile components like nitrides and hydrides although details will depend on the proporties of the specific system.  

\medskip
\noindent\textbf{Experimental Section} \par 

A 400\,nm thick PCMO film was grown on a commercial (100) oriented STO substrate by ion beam sputtering from a single target of composition ($x=0.1$). Deposition parameters were: $p_\text{Ar}=3\times10^{-2}$\,Pa (beam neutralizer), $p_\text{Xe}=1\times10^{-2}$\,Pa (sputter gas), and $p_\text{O}=1.6\times10^{-2}$\,Pa (film oxidation). The deposition temperature was set to 820$^\circ$\,C using a Tectra boron nitride heater which results in approx. 720$^\circ$\,C at the substrate surface. Subsequently, the film was annealed in air for 20h at 900$^\circ$\,C with a ramping speed of 100$^\circ$\,C per hour.

TEM lamellas were extracted using an FEI Nova NanoLab Dual Beam focused ion beam and mounted on first generation four-contact DENSsolutions heating chips. The final thinning step was done using an acceleration voltage of 5\,kV. TEM experiments were conducted in an FEI Titan ETEM G2 80-300 operated at 300\,kV using a DENSsolutions Lightning D7+ holder and a semi-convergence angle of 0.8\,mrad. NBED patterns were recorded with a self-written DigitalMicrograph plugin controlling the beam position and reading out an UltraScan 1000XP camera binned to 256$\times$256 pixels.

In order to extract the lattice parameters from resulting NBED patterns, their auto-correlation is calculated in a first step. Subsequently, a two-di\-men\-sio\-nal reciprocal lattice is fitted up to the second order to the positions of the local maxima at 
\begin{equation}
\mathbf{k}(i,j)=(k_x(i,j),k_y(i,j)),\; -2 \leq i,j \leq 2
\end{equation}
closest to the multiple orders of an initial guess 
$(i\mathbf{b}_{10}+j\mathbf{b}_{20})$
by solving the linear optimization problem:
\begin{equation}
\min_{\mathbf{b}_1,\mathbf{b}_2}\sum_{i,j}I(i,j)^2\left(\mathbf{k}(i,j)-i\mathbf{b}_{1}-j\mathbf{b}_{2}\right)^2
\end{equation}
Here, $I(i,j)$ corresponds to the intensity of the local maximum at $\mathbf{k}(i,j)$ and serves as a weighting factor. Please note that this strategy of lattice parameter extraction is inspired by and thus very similar to those of previous reports.\textsuperscript{\cite{LookInOphus,baumann}}
Throughout this paper, the manually estimated (110) and (001) reflections of the STO substrate at room temperature were used as initial guesses $\mathbf{b}_{10}$ and $\mathbf{b}_{20}$. In the orthorhombic Pbnm unit cell of PCMO, these reflections translate approximately to (100)/(010) and (002). The moduli of real space lattice parameters were calculated by taking the inverse of the moduli of the reciprocal lattice parameters.

\medskip
\noindent\textbf{Acknowledgements} \par 
The project was funded by the Deutsche Forschungsgemeinschaft (DFG, German Research Foundation) -- 217133147/SFB 1073, projects B02, Z02.

\medskip
\noindent\textbf{Author Contributions} \par 
The design of experiments and their interpretation was the result of fruitful discussions including all authors; T.M. extracted the TEM lamellas from a thin film sample grown by B.K. and conducted the TEM experiments together with V.R. and B.K.; The data was analyzed by T.M. under revision of M.S.; The manuscript was written by T.M. under revision of M.S. and C.J.; All authors read and agreed on the written paper;

\medskip
\noindent\textbf{Conflict of Interest} \par 

The authors declare no conflict of interest.

\bibliographystyle{apsrev}

\end{document}